\documentclass[twocolumn,aps,prl,floatfix,superscriptaddress]{revtex4}

\usepackage{amsmath}    % need for subequations
\usepackage{graphicx}   % need for figures
\usepackage{epsfig}
\usepackage{verbatim}   % useful for program listings
\usepackage{color}      % use if color is used in text
\usepackage{subfigure}  % use for side-by-side figures
\usepackage{hyperref}   % use for hypertext links, including those to external documents and URLs
\usepackage{bm}
\usepackage{amsfonts}
\usepackage{amssymb}

\def\buk{{\bm u}_{\bm k}}

\def\upp{u^+_{\bm p}}
\def\uqm{u^-_{\bm q}}
\def\uqm{u^-_{\bm q}}

\def\ukp{u^+_{\bm k}}
\def\ukm{u^-_{\bm k}}

\def\bk{{\bm k}}

\def\bq{{\bm q}}
\def\bu{{\bm u}}
\def\bx{{\bm x}}
\def\bz{{\bm z}}
\def\bw{{\bm \omega}}
\def\hm{{\bm h}^-_{\bm k}}
\def\hp{{\bm h}^+_{\bm k}}
\def\hpm{{\bm h}^\pm_{\bm k}}

\usepackage{color}

\begin{document}
\title{Role of helicity for large- and small-scale turbulent fluctuations\footnote{Postprint version of the manuscript published in Phys. Rev. E {\bf 92}, 051002 (R) (2015).}} 
\author{Ganapati Sahoo\footnote{ganapati.sahoo@gmail.com}, Fabio Bonaccorso, and Luca Biferale}
\affiliation{Department of Physics \& INFN, University of Rome Tor Vergata, Via della Ricerca Scientifica 1, 00133 Rome, Italy.}
\date{\today}

\begin{abstract}
Effects of the helicity  on the dynamics of turbulent flows  are
investigated. The aim is to disentangle the role of helicity in fixing the
direction, the intensity, and  the fluctuations of the energy transfer across
the inertial range of scales. We introduce an external parameter $\alpha$
that controls the mismatch between the number of positive and negative
helically polarized Fourier modes. We present direct numerical
simulations of Navier-Stokes equations from the fully symmetrical case,
$\alpha=0$, to the fully asymmetrical case, $\alpha=1$, when only helical modes
of one sign survive. We found a singular dependency  of the direction of the
energy cascade on $\alpha$, measuring a positive forward flux as soon as only a
few modes with different helical polarities are present. 
Small-scale fluctuations are also strongly sensitive to the degree of mode-reduction,
leading to a vanishing intermittency already for values of $\alpha \sim 0.1$.
 If the analysis is restricted to sets of modes with the same helicity sign, intermittency is 
vanishing for the modes belonging to the  minority set, and it is close to that measured on the original
 Navier-Stokes equations  for the other set. 
\end{abstract}

\maketitle

The direction of the energy transfer in a turbulent flow is believed to be
determined by the combined effects of all inviscid invariants which depend on
the embedding dimensionality and/or on the coupling with external fields, as in
conducting or buoyant systems \cite{biskamp,frisch,lohse_arfm}. For fully
homogeneous and isotropic turbulence (HIT) in two dimensions, the presence of
two positive-definite invariants, energy and enstrophy, leads to
a split regime with energy flowing towards  large scales
(inverse cascade) and enstrophy to small scales (forward cascade)
\cite{kraichnan_2d,boff_2d,tabeling,2dreview,ecke,prl_massimo}. 
Three-dimensional (3D) Navier-Stokes equations (NSEs) possess two inviscid
invariants; energy and helicity, the scalar product of velocity and vorticity
\cite{moffatt69,moffatt92,brissaud}. Different from the energy, helicity
is observed to be preserved by some dissipative events, such as antiparallel
vortex reconnection \cite{Laing,Sheeler}, and it is not positive definite.  As a
result, it is not possible to predict the direction of the energy and helicity
transfers from fundamental arguments. 
Numerical simulations, phenomenological
arguments, dynamical models, closures, and comparison with the inviscid
Gibbs-like equilibrium distribution suggest that both energy and helicity develop
a forward cascade in HIT
\cite{brissaud,kraichnan,waleffe,chen,chen2,ditlevsen,biferaleh,EDQNM}.  On the
other hand, it is well known that the external mechanisms such as rotation
\cite{mininni,deusebio}, confinement \cite{celani,xia}, shear \cite{dubrulle} or
coupling with the magnetic field \cite{brandenburg} might revert the direction
of the energy cascade.
Strikingly enough, such a reversal of the flux has been predicted and observed
also in 3D HIT with explicit breaking of parity invariance, i.e., by
restricting the dynamics to a subset of Fourier modes such that the helicity
becomes sign definite \cite{biferale2012,biferale2013,herbert}, suggesting
that inverse energy transfer events are much broader than previously thought
and they are potentially present in all flows in nature. 
Another not understood crucial
aspect of fully developed turbulence is {\it intermittency}, the tendency of
the flow  to develop more and more non-Gaussian velocity fluctuation at smaller
and smaller scales.  It is not known which key  degrees of freedom
are responsible for such a phenomenon and whether it is crucial to
keep all interactions among different scales to preserve it.  Consequently, we
cannot model it and we cannot predict its degree of universality, i.e.,
independence of the large-scale configuration. There exist models based on
strong mode reduction (shell models \cite{biferale_shellmodel}) which  preserve
many of the intermittent properties of the original Navier-Stokes equations.
On the other hand,  an increase or decrease of intermittency is observed in
numerical simulations of the original 3D equations with a modulation of
local or nonlocal Fourier interactions \cite{laval}.  In this Rapid Communication we
systematically investigate the effects of  helical mode reduction in 3D NSEs.
The aim is threefold: We want to explain the role played by helicity in
fixing (i) the direction, (ii) the intensity, and (iii) the intermittency of
the energy transfer mechanism.

The key tool used is based on a suitable projection of the NSEs allowing one to
disentangle, {\em triad by triad}, the properties of the  energy transfer as a
function of the percentage of negative helically polarized modes kept in the
simulation. The existence of a control parameter is crucial to address the
problem in a quantitative way, tailoring  the degrees of freedom retained and
removed, without any modeling.  We start with the helical decomposition
\cite{waleffe,sagaut}
of the velocity field $\bu(\bx)$, expanded in a Fourier series as 
  $\buk  = \ukp \hp + \ukm \hm$,
where $\hpm$ are the eigenvectors of the curl, i.e., $i {\bk} \times \hpm = \pm k \hpm$. 
 We choose $\hpm = \hat{\nu}_{\bm k} \times
\hat{k} \pm i \hat{\nu}_{\bm k}$, where $ \hat{\nu}_{\bm k}$ is a
unit vector orthogonal to ${\bk}$, satisfying the condition
$\hat{\nu}_{\bm k} = - \hat{\nu}_{-\bk}$, e.g., $\hat{\nu}_{\bk} = {\bz} \times {\bk}
/|| {\bz} \times {\bk} ||$, with any arbitrary vector ${\bz}$.
 In terms of such an {\it exact} decomposition of each Fourier mode, 
the total energy, $E = \int d^3 x \, |\bu(\bx)|^2$, and the 
total helicity, $H = \int d^3 x \, \bu(\bx) \cdot \bw(\bx)$, are written as
\begin{align} 
    E = \sum_{\bk} |\ukp|^2 + |\ukm|^2,\quad H = \sum_{\bk} k(|\ukp|^2 - |\ukm|^2),
\end{align} 
where $\bw$ is the vorticity (see also Refs. \cite{morinishi,chen} for other previous applications of the same decomposition). The nonlinear term of the NSE can be
then decomposed in terms of the helical content of the complex amplitudes,
$u^{s_\bk}_{\bk}$ with $s_\bk = \pm $ (see Ref. \cite{waleffe}). 
We consider the dynamics of an incompressible flow (${\bm \nabla} \cdot {\bf
\bm u} = 0$) determined by the decimated NSE in which a
fraction $\alpha$ of the negative helical modes has been switched off
\cite{footnote}. We introduce the projector on
positive/negative helical modes as
\begin{align}
  \label{eq:proj}
  {\mathcal P}^\pm_{\bk} \equiv \frac {\hpm \otimes \overline{\hpm}} {\overline{\hpm} \cdot \hpm},
\end{align}
where $\overline{\bullet}$ denotes the complex conjugate. We define an
operator ${D}^\alpha$ that projects each wavenumber with a probability $ 0
\le \alpha \le 1$,
\begin{equation} 
  \label{eq:projv}
  {\bu}^\alpha(\bx) \equiv D^{\alpha} {\bu}(\bx) \equiv \sum_{\bk} e^{i{\bm k}\bx}\, {\mathcal D}^\alpha_{\bk} {\buk},  
\end{equation}
where 
%\begin{equation}
\({\mathcal D}^\alpha_{\bk}  \equiv (1-\gamma^\alpha_{\bk}) + \gamma^\alpha_{\bk} {\mathcal P}^+_{\bk}\)
%\end{equation}
and the random numbers are $\gamma^\alpha_{\bk}=1$  with probability $\alpha$ or $\gamma^\alpha_{\bk}=0$ with probability $1-\alpha$. 
The $\alpha$-decimated Navier-Stokes equations ($\alpha$-NSE) are
\begin{equation} 
  \label{eq:ns+++}
  \partial_t \bu^\alpha = D^{\alpha}[- \bu^\alpha \cdot {\bm \nabla} \bu^\alpha -{\bm \nabla} p^\alpha] 
  +\nu \Delta \bu^\alpha, 
\end{equation}
where $\nu$ is the viscosity and $p$ is the pressure. 
Notice that the nonlinear terms on the right-hand side of (\ref{eq:ns+++}) are further projected by
${ D}^\alpha$ in order to enforce the dynamics  on the
selected set of modes for all times. Despite the fact 
that the $\alpha$-NSE break the Lagrangian properties of the nonlinear terms \cite{moffatt14},
 both  energy, $ E = \sum_{\bk} ( |\ukp|^2 + (1-\gamma_{\bk})|\ukm|^2)$, and helicity, 
 $H = \sum_{\bk} k( |\ukp|^2 - (1-\gamma_{\bk})|\ukm|^2), \label{eq:halpha}$
are still  invariants in the inviscid limit of (\ref{eq:ns+++}), as in any Galerkin truncation.
We can then  identify  two extreme cases: When $\alpha=0$, we recover the
original NSE, and when  $\alpha=1$, helicity becomes a coercive  quantity with a
definite sign. It has been recently shown~\cite{biferale2012,biferale2013} that,
in the latter case, the dynamics of (\ref{eq:ns+++}) develops a double cascade
characterized by an inverse energy transfer with a Kolmogorov spectrum $E(k) \sim
k^{-5/3}$ for wavenumbers smaller than the forcing scale,  $k \ll k_f$, and a
direct helicity cascade with a $k^{-7/3}$ spectrum for $k \gg k_f$. Here 
we address the questions: Does there
exist a critical value $\alpha_c$ where the direction of the mean energy
transfer suddenly reverses as observed in two-dimensional hydromagnetic systems
when changing the forcing mechanisms \cite{alexakis}, or does the helicity play a
singular role?  Are a few modes with opposite helical sign enough to transfer
energy to small scales ($\alpha_c \rightarrow 1$), as suggested in
Ref. \cite{herbert} from considerations based on absolute equilibrium?  What happens
to small-scale intermittency in the forward cascade  regime?  Does it depend
on the amount of negative/positive helical modes retained? Is the residual
small-scale vorticity mainly helical?  In order to answer these key
questions, we performed a series of numerical simulations at changing
$\alpha$ with a fully dealiased, pseudospectral code at resolution up to
$1024^3$  on a triply periodic cubic domain of size $L=2\pi$.  The flow is
sustained by a random Gaussian forcing with \[\langle f_i(\bk,t) f_j(\bq,t')
\rangle = F(k) \delta(\bk-\bq) \delta(t-t') Q_{ij}(\bk),\] where $Q_{ij}(\bk)$
is a projector assuring incompressibility and $F(k)=F_0k^{-3}$. $F_0$ is nonzero only for $ |k|
\in [k_{\rm min}:k_{\rm max}]$. See Table.~\ref{table1} for details of the
simulations. A different forcing process based on a second-order
Ornstein-Uhlenbeck process has also been implemented for some simulations with
the same results (not shown).  In all cases we have used a fully helical
forcing with projection only on $\hp$ in order to ensure a maximal helicity
injection rate $h$, independent of the degree of decimation $\alpha$ of
the negative helical modes.
%%%%%%%%%%%%%%%%%%%%% table 1 %%%%%%%%%%%%%%%%%%%%%%%%%%%%%%%%
\begin{table}
  \begin{center}
  \begin{tabular}{| c | c | c | c | c | c | c |}
  \hline
   RUN   &  1-8  & 9-13 & 14-19 & 20 & 21 & 22 \\
  \hline
   $N$ & $256$  & $256$   & $512$ & $512$ & $1024$ & $1024$ \\ 
   $k_f$ &  $[1,3]$ & $[1,3]$ & $[1,2]$ & $[42,50]$ & $[10,12]$ & $[1,2]$ \\ 
   $\alpha$ & $0-0.999$ & $0.1-0.9$ & $0-0.9999$ & $1.0$ & $1.0$ & $0$ \\ 
   $F_0$ & $1.0$ & $1.0$ & $1.0$ & $10.0$ & $1.0$ & $1.0$ \\ 
  \hline
  \end{tabular}
  \end{center}
  \caption{$N$: Number of collocation points along each axis. $k_f$: forced
wavenumbers. $F_0$: Forcing amplitude. RUN 1-8: Decimation of only negative helical modes with
probability in the range $\alpha \in [0:0.9999]$. RUN 9-13: Same
 of RUN 1-8  but with either positive or negative helical modes (with
$50\%$ probability) removed.  RUN 14-19: Similar to RUN 1-8 at higher resolutions. RUN 20-21:
Forced at small scales. RUN 22: Same as
RUN 1 at higher resolution.}
  \label{table1}
\end{table}
%%%%%%%%%%%%%%%%%%%%%%%%%%%%%%%%%%%%%%%%%%%%%%%%%%%%%%%%%%%%%%
%%%%%%%%%%%%%%%%%%%   fig 1   %%%%%%%%%%%%%%%%%%%%%%%%%%%%%%%%
\begin{figure}[!h]
  \includegraphics[scale=0.65]{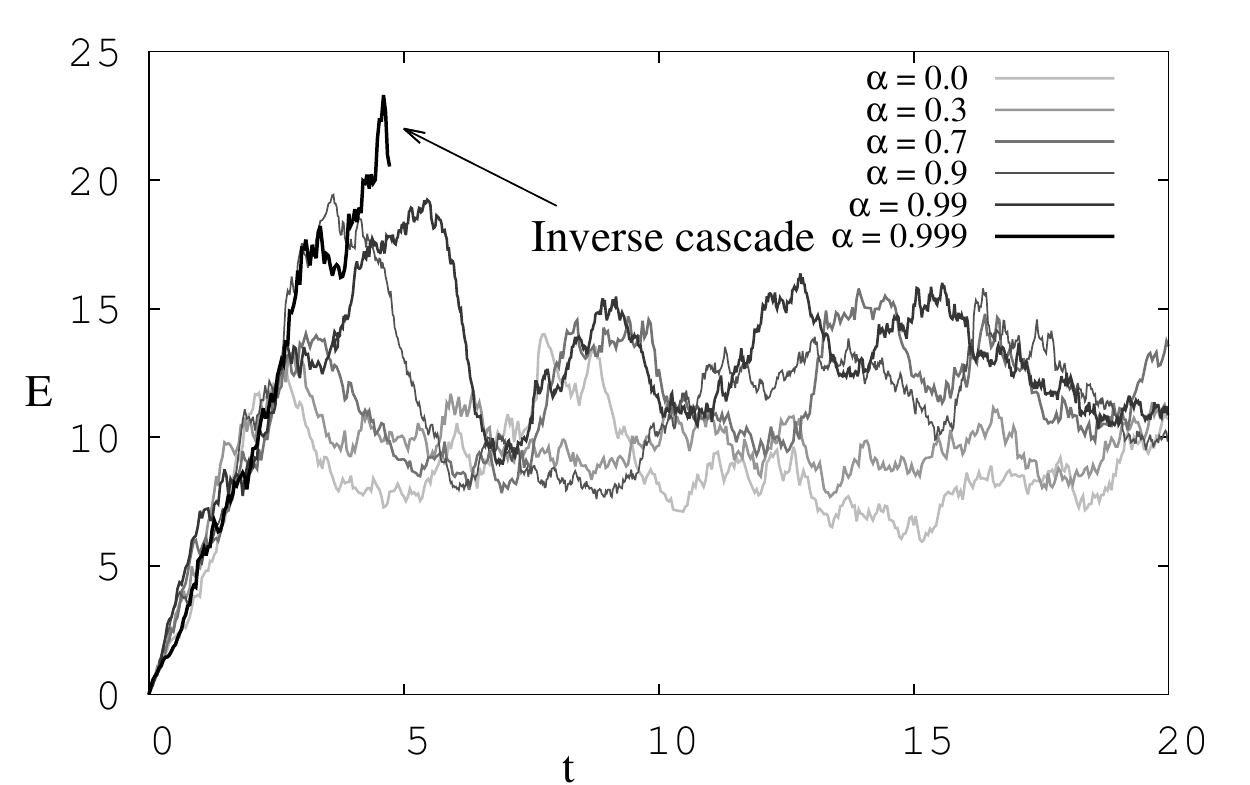}
  \caption{Evolution of energy at varying $\alpha$.  All simulations
reach a stationary state, except for the case at $\alpha=0.999$, where a constant
increase of energy is observed, signaling the existence of a stable
inverse energy transfer. }
  \label{fig:1}
\end{figure}
%%%%%%%%%%%%%%%%%%%%%%%%%%%%%%%%%%%%%%%%%%%%%%%%%%%%%%%%%%%%%%
%%%%%%%%%%%%%%%%%%%   fig 2   %%%%%%%%%%%%%%%%%%%%%%%%%%%%%%%%
\begin{figure}[!htb]
  \includegraphics[scale=0.65]{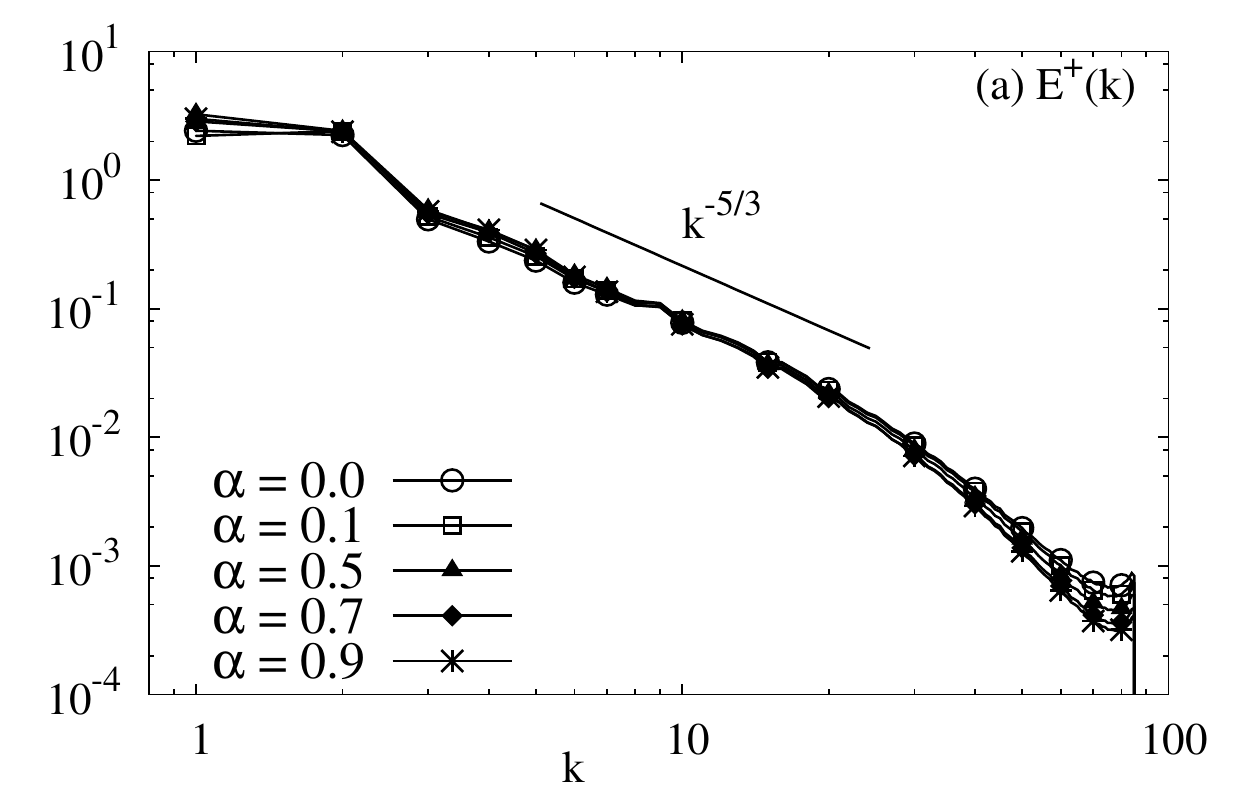}\\
  \includegraphics[scale=0.65]{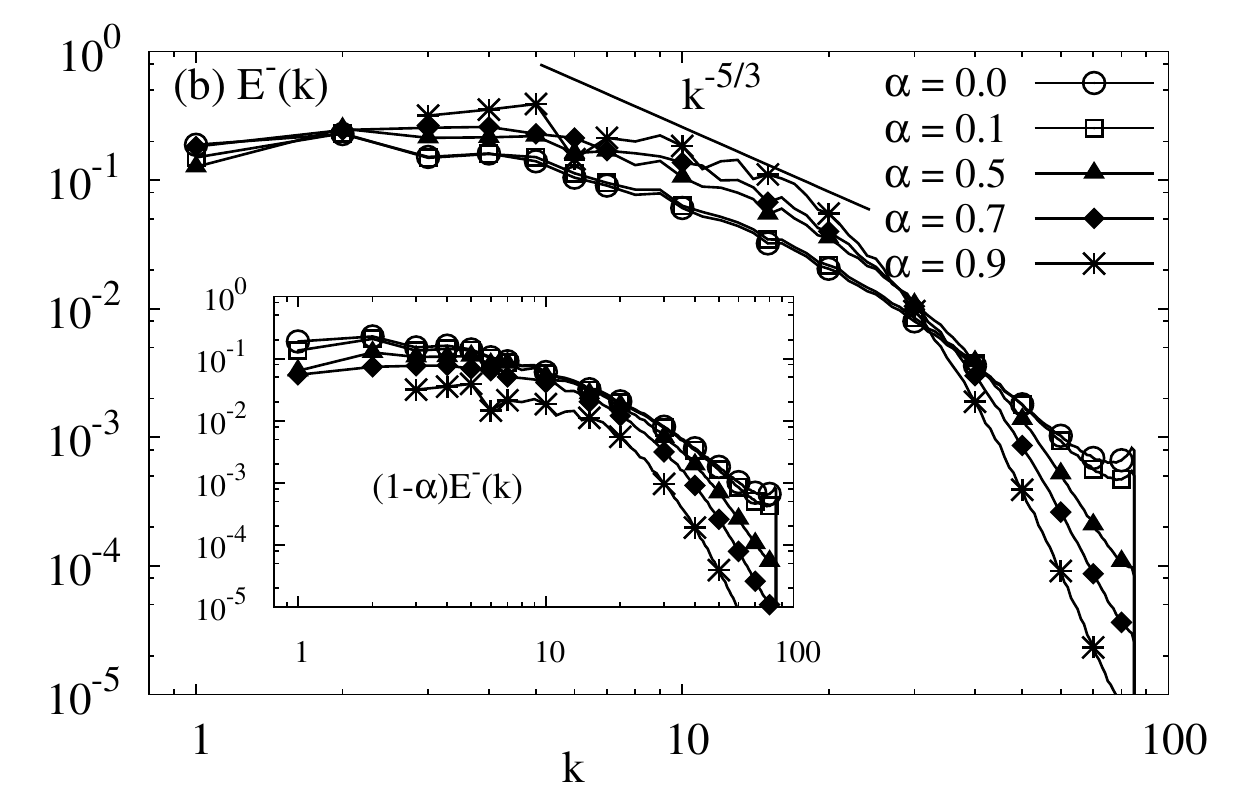}\\
  \includegraphics[scale=0.65]{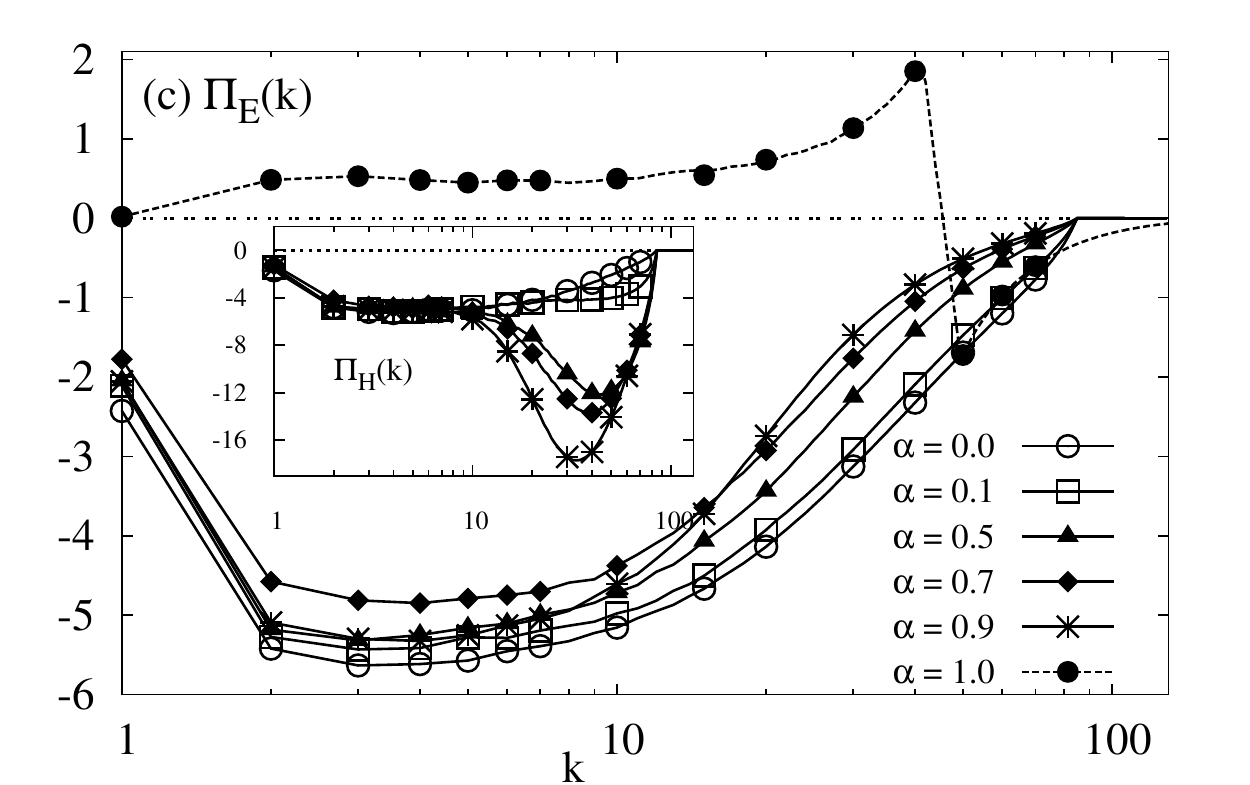}
  \caption{(a) Log-log plot of $E^+(k) = \sum_{|\bk|=k}|\ukp|^2$ vs $k$ at
changing $\alpha$. (b) Log-log plot of $E^-(k) = \sum_{|\bk|=k}(1-\gamma_\bk)
|\ukm|^2$ vs $k$ at changing $\alpha$. Inset: Rescaled $E^-(k)$ with factor $(1-\alpha)$.
(c) Semilog plot of flux of energy. Inset: Flux of helicity, at changing
$\alpha$.}
  \label{fig:2}
\end{figure}
%%%%%%%%%%%%%%%%%%%%%%%%%%%%%%%%%%%%%%%%%%%%%%%%%%%%%%%%%%%%%%

\noindent {\sf Energy transfer.} We start from  the spectral properties of the system following
Ref. \cite{chen}.  We define the total spectra restricted to the positive/negative
helical modes as $E^{+}(k) = \sum_{|\bk|=k} |\ukp|^2, E^{-}(k) = \sum_{|\bk|=k}
(1-\gamma_\bk)|\ukm|^2 $, and the corresponding quantity for the helicity,
$H^{\pm}(k) =  k\,E^{\pm}(k) $. In the case when both energy and helicity are
transferred forward with a rate $\varepsilon$ and $h$, respectively, we expect
the usual Kolmogorov 1941 scaling (K41) for both energy and helicity spectra
\cite{brissaud,chen},
\begin{align*}
E(k) \sim C_E\varepsilon^{2/3}k^{-5/3},\quad H(k) \sim C_H h\varepsilon^{-1/3} k^{-5/3},
\end{align*}
 which reflects in the scaling for each component as
\begin{equation}
E^{\pm}(k) =   \varepsilon^{2/3}k^{-5/3}[1 \pm Ch(\varepsilon k)^{-1}],
\end{equation}  
where $C=C_H/C_E$. In Fig.~\ref{fig:1} we show the time evolution of the total
energy $E$ starting from a null configuration at $t=0$ when varying the degree of
decimation from $\alpha=0$, for the nondecimated NS case, to $\alpha \sim 1$.
We notice first that  the  time needed to develop the initial release of energy
becomes longer with increasing $\alpha$, and that the oscillations around the
stationary regime are also larger when $\alpha \sim 1$.  The most striking
phenomenon is that even for a very high decimation of negative helical modes,
$\alpha \sim 1$, the system is able to reach  a stationary state by transferring
energy to the small scales. In other words, it is enough to have very few
negative helical modes to develop a  stable and stationary positive energy
flux.  This is  quantified in Fig.~\ref{fig:2}, where we separately plot the
spectra for the two helical components for various $\alpha$. The spectrum for
the positive helical modes [Fig.~\ref{fig:2}(a)] is almost unchanged and
independent of $\alpha$ with a clearly developed $k^{-5/3}$ slope, whereas the
spectrum for the negative helical modes [Fig.~\ref{fig:2}(b)] tends to react back
and become more and more energetic as $\alpha$ increases; this can be explained
by looking at the behavior of the energy flux.  In Fig.~\ref{fig:2}(c) we show
that the  energy flux is constant and independent of $\alpha$ for all $\alpha
<1$---it reverts only for $\alpha \sim 1$. The surprising efficiency of the
nonlinear transfer suggests that  helicity plays a singular role in turbulence:
A tiny mixture of positive and negative helical modes axccross all scales is
enough to sustain a forward energy. This can be reconducted to the role of 
the triads with two high-wavenumber modes of opposite
helicity~\cite{waleffe}. If this is the case, the most important triads must 
have one negative and two positive helical modes:
\begin{equation}\label{eq:triad}
S(k|p,q) = \langle (\bk\cdot \uqm)(\ukp \cdot \upp)\rangle + \langle (\bk
\cdot \upp) (  \ukp \cdot \uqm)\rangle.
\end{equation}
They are present with probability $\propto (1-\alpha)$ while triads
 with two negative  helical modes, exist with 
probability $\propto (1-\alpha)^2$. To keep the above triadic correlation constant at decreasing $\alpha$, we must have:
$\ukm \rightarrow \ukm/(1-\alpha)$ and therefore:
    $E^-(k) = \sum_{|\bk|=k} (1-\gamma_\bk) |\ukm|^2 
\rightarrow E^-(k)/(1-\alpha).$ 
This prediction is shown to be well realized  in the
inset of Fig.~\ref{fig:2}(b), where we show that rescaling  $E^-(k)$ by a
factor $(1-\alpha)$ leads to a good overlap, except for $\alpha \sim 1$,
where the fluctuations due to the onset of the inverse energy transfer becomes
very large and the above  argument possibly breaks down. Thus, negative helical modes
act as a {\it bridge} for the energy transfer; they receive
energy from the large-scale positive modes and release it to the
small-scale positive modes; the fewer there are, the more intense their amplitude
must be to do it efficiently.  When negative 
helical modes become too rare or absent, i.e., for $\alpha\sim 1$, this bridging is no
longer possible and the energy flows upscale \cite{biferale2012}. Helicity
plays the role of a passive catalyst in the energy transfer. 
Proving the existence of
a unique $\alpha_c$ for the inversion of the energy transfer could be extremely
hard and it may not be crucial. The observed value is so close to unity that it
might also be dependent on the realization of $\gamma_{\bm k}$ and/or on the
Reynolds numbers. This issue is left for a more detailed analysis in a future
work. 

%%%%%%%%%%%%%%%%%%   fig 3   %%%%%%%%%%%%%%%%%%%%%%%%%%%%%%%%
%%%%%%%%%%%%%%%%%%%%%%%%%%%%%%%%%%%%%%%%%%%%%%%%%%%%%%%%%%%%%%
\begin{figure}[!htb]
  \includegraphics[scale=0.9]{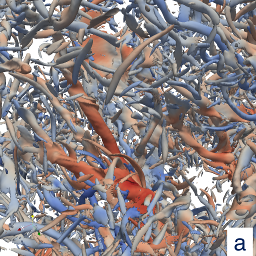}\\
  \includegraphics[scale=0.9]{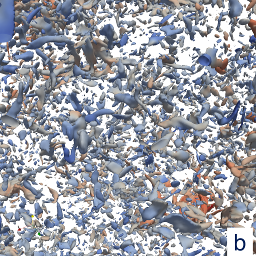}
  \caption{ (Color online)  Iso-vorticity surfaces for (a) $\alpha = 0$, (b)
$\alpha= 0.5$. The color palette is proportional to the  intensity of the 
helicity, red for high positive values ($\sim 10^3$) to blue
for high negative values ($\sim -10^3$).}
  \label{fig:3}
\end{figure}
%%%%%%%%%%%%%%%%%%%%%%%%%%%%%%%%%%%%%%%%%%%%%%%%%%%%%%%%%%%%%%
\noindent {\sf Intermittency.} The second important problem addressed concerns
intermittency, the presence of strong non-Gaussian fluctuations at small
scales, usually interpreted as a build up of instabilities in the
vortex-stretching mechanisms.  Here, we want to understand how intermittency
changes under helical mode reduction. A visual inspection of the
vorticity field, in Fig.~\ref{fig:3}, shows a strong depletion of filament-like
structures, of the standard 3D NSE [Fig.~\ref{fig:3}(a)] with decimation of the
negative helical modes, shown, e.g., for $\alpha=0.5$ [Fig.~\ref{fig:3}(b)]. In
order to quantitatively assess the degree of intermittency, we
focus on the so-called structure functions (SFs), $ S^{(p)}(r)= {\langle (\delta_{r} u^\alpha)^p \rangle}$,
 based on the $p$ moments of the transverse velocity increments $\delta_{r} u^\alpha = u^\alpha_
y(x+r)-u^\alpha_y(x)$ as a function of the separation scale (the selection of
the $x-y$ components is arbitrary because of isotropy).
%\begin{align*}
%S^{(p)}(r)= {\langle (\delta_{r} u^\alpha)^p \rangle}.
%\end{align*}}
%%%%%%%%%%%%%%%%%%fig 4   %%%%%%%%%%%%%%%%%%%%%%%%%%%%%%%%
%%%%%%%%%%%%%%%%%%%%%%%%%%%%%%%%%%%%%%%%%%%%%%%%%%%%%%%%%%%%%%
\begin{figure}[!htb] 
\includegraphics[scale=0.67]{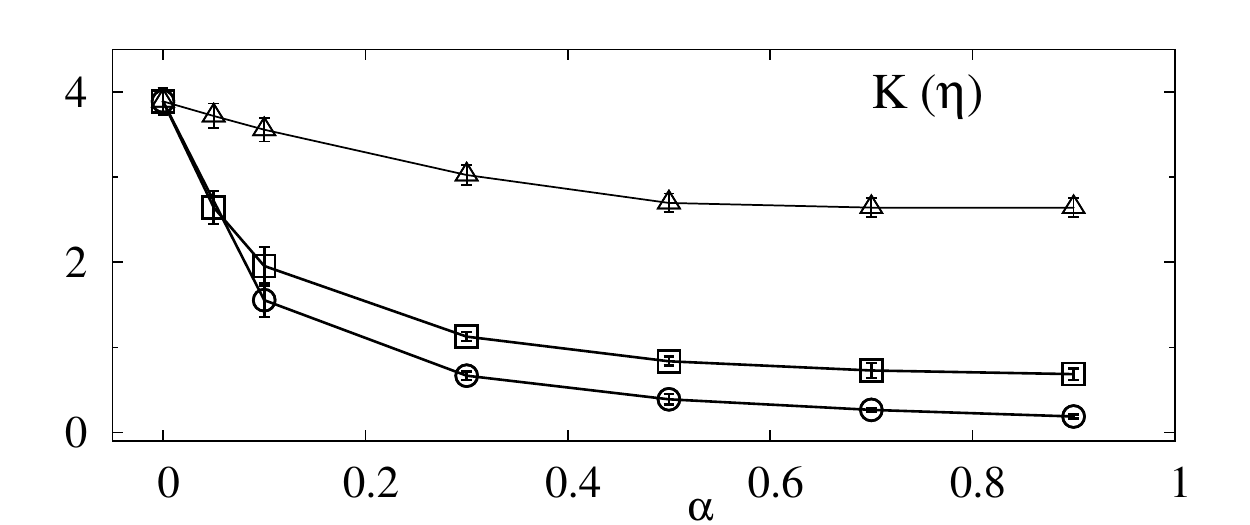}\\
\hspace{-0.5cm}\includegraphics[scale=0.7]{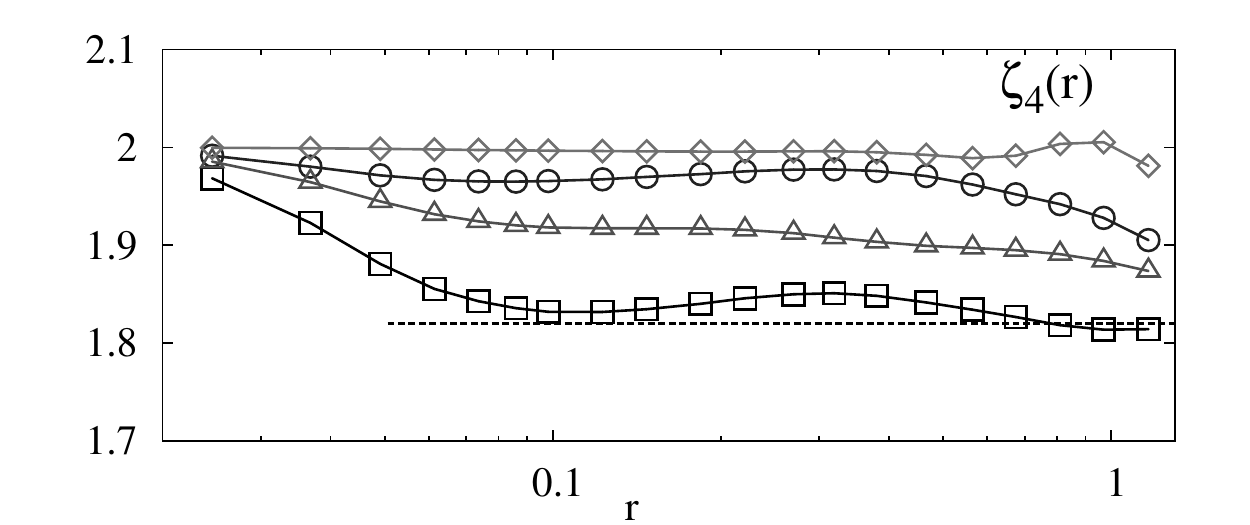}
\caption{ Top panel: Excess kurtosis of transverse SF for $r=\eta$ at changing
$\alpha$: ($\square$): decimation  of negative helical modes only;
($\bigcirc$): decimation of either positive or negative helical modes with
$50\%$ probability; ($\triangle$): {\em a posteriori} decimation of negative
helical modes from a velocity field of standard nondecimated NSEs.  Bottom:
Logarithmic local slopes for fourth order moment. $\alpha=0$, i.e., the full NSE
($\square$). $\alpha=0.5$ : The whole field ($\bigcirc$) or its $\hp$ ($\triangle$)
and $\hm$ ($\Diamond$) components. The horizontal line represents the standard
intermittent correction as predicted from Ref. \cite{she}.}
\label{fig:4} 
\end{figure}
%%%%%%%%%%%%%%%%%%%%%%%%%%%%%%%%%%%%%%%%%%%%%%%%%%%%%%%%%%%%%%
In Fig.~\ref{fig:4} we show
(i) the local slopes of the relative scaling of fourth order SF with respect to the second order, 
$ \zeta^{(4)}(r) = d \, log[S^{(4)}(r)]/d\, log[S^{(2)}(r)]$ for the case $\alpha=0.5$, and (ii) the value of the  excess
kurtosis  for $r=\eta $, $K(\eta) = S^{(4)}(\eta)/[S^{(2)}(\eta)]^2$, at changing
$\alpha$.
In the top panel, concerning the kurtosis, we found that intermittency is highly sensitive to
$\alpha$ decimation; it is enough to remove a small fraction of negative helical modes 
  from the dynamics to strongly deplete the non-Gaussian character. In the same panel 
we show  also the results of another numerical experiment,
where we  repeated the measurements in a set of simulations (RUN 9-13) with
{\em random} decimation; this time either a positive {\em or} a negative
helical mode is decimated with a  probability $\alpha$. The reduction in
the intensity of intermittency is comparable with the previous case.
To further investigate  the role of dynamic helical mode-reduction, we
performed a projection  {\it aposteriori}, applying the operator  $D^{\alpha}$  
to the velocity field obtained from
a fully resolved non-decimated NSE ($\alpha=0$).  In this case, intermittency
remains almost unchanged, independently of $\alpha$, suggesting that only
the dynamical mode reduction is crucial to deplete the vortex-stretching mechanism.
In the bottom panel  we show the results concerning the local scaling 
exponent $\zeta^{(4)}(r)$ for $\alpha=0.5$, and we compare it with the original 
NS case $(\alpha=0)$ and with two other important measurements obtained by 
taking the velocity configurations  dynamically generated with $\alpha=0.5$ 
and projecting them on their positive or negative helical components, i.e., by applying the 
projector $ {\mathcal P}^+_{\bk}$ or ${\mathcal P}^-_{\bk}$ on all modes. 
Doing that, we observe intermittency for the projection on the majority component $\hp$ and
a vanishingly small correction for the projection on the minority helical component $\hm$. In summary,
we can conclude that intermittency as measured from real-space 
velocity configurations is the results of a highly 
non-trivial and entangled correlation among a subset of key modes in Fourier space.
It is hidden in the correlation among all Fourier modes with given helical components.

\noindent {\sf Conclusion.} We have highlighted and quantified the singular
role played by helical Fourier modes in the energy flux reversal, showing
that a forward transfer is always preferred as soon as  a very small percentage
of modes with opposite helicity are present. These findings suggest the
possibility to check \emph{a posteriori} on direct numerical simulations and experiments of
strongly rotating flows, of flows under vertical confinement or under strong
shear the role played by different helical triadic interactions in driving the energy transfer
forward or backward.  Intermittency measured on 
real-space fields is fragile and strongly dependent on the mode-reduction
protocol,  suggesting that its origins  must rely on highly nontrivial
correlations among helical and nonhelical fluctuations. In particular, it is
apparently key to have all modes with a given helical component.  Another key
factor might be to keep  local or nonlocal interactions in Fourier space, as
also suggested by numerical experiments based on a scale-dependent
mode-reduction scheme \cite{laval}.
\begin{acknowledgments}
We acknowledge funding from the European Research Council under the European
Union's Seventh Framework Programme, ERC Grant Agreement No 339032.
\end{acknowledgments}


\begin{thebibliography}{99}
\bibitem{footnote} The choice of decimating negative mode is purely arbitrary nothing would change by decimating positive modes.
\end{thebibliography}
\end{document}